\documentclass[pre,twocolumn,superscriptaddress,showpacs,floatfix ]{revtex4}
\usepackage{graphicx}
\usepackage{dcolumn}
\usepackage{bm}

\begin{document}
\newcommand{\bb}{\begin{equation}}
\newcommand{\ee}{\end{equation}}
\newcommand{\dd}{{\rm d}}
\newcommand{\dr}{{\rm d}{\mathbf r}}
\newcommand{\dR}{{\rm d}{\mathbf R}}
\newcommand{\nr}{n({\mathbf r})}
\newcommand{\rhor}{\rho({\mathbf R})}
\newcommand{\rr}{{\mathbf r}}
\newcommand{\RR}{{\mathbf R}}
\newcommand{\dz}{{\rm d}z}

\title{Density functional approach for inhomogeneous star polymers} 
\author{A. Malijevsk\'y}
\affiliation{Institute of Theoretical Physics,
Faculty of Mathematics and Physics, Charles
University Prague, V Hole$\check{s}$ovi$\check{c}$k\'ach 2, 180 00 Praha 8,
Czech Republic}
\author{P. Bryk}
\email{pawel@paco.umcs.lublin.pl}
\affiliation{ Department for the Modeling of Physico-Chemical Processes,
MCS University,  20-031 Lublin, Poland}
\author{S. Soko\l owski}
\affiliation{ Department for the Modeling of Physico-Chemical Processes,
MCS University,  20-031 Lublin, Poland}
\date{\today}
\begin{abstract}
\noindent We propose microscopic density functional theory   
for inhomogeneous star polymers. Our approach is based on fundamental
measure theory for hard spheres, and on Wertheim's  first- and
second-order perturbation theory for the interparticle
connectivity. For simplicity we consider a model in which all the arms
are of the same length, but our approach can be easily extended
to the case of stars with arms of arbitrary lengths.

\end{abstract}
\maketitle

\newpage

It has been demonstrated 
that density functional theory
(DFT) is a versatile and powerful tool to represent
the structural and thermodynamic properties of polymeric fluids
\cite{Likos,Mayer,Woodward91,Yeti95,Yeti,Yu}.
Taking into account the level of physical model,
DFT's of polymers can be divided into two main categories.
The first category
\cite{Likos,Mayer} involves the so-called coarse-graining procedure
\cite{Louis00}, in which the degrees of freedom of monomers building the
polymer coils are integrated out. The resulting effective pairwise potential
between the centers of masses of two molecules is then used in further
investigations \cite{Archer02}. An 
advantage of
these models emerges
from the possibility of application of
theories of simple fluids to describe
the polymers. 

It is obvious that coarse-grained models lose some
information, e.g. a possibility of evaluation
of correlation functions between
the monomers. From this point of view, models of the second category
\cite{Yeti,Yu,Yu1,Bryk04,Forsman:02:1,Forsman:04:2}, which explicitly treat
the microscopic structure of
polymers seem to be superior. Several microscopic DFT approaches 
for inhomogeneous chain polymers have been proposed in the literature. 
Woodward and Yethiraj
\cite{Woodward91,Yeti95,Yeti} developed a theory that combines  weighted
density approximation, known from theories of simple fluids, with
single-chain Monte Carlo simulations. An alternative DFT of inhomogeneous
polymer solutions was formulated by Forsman, Woodward and
Freasier\cite{Forsman:02:1}. Their theory
is based on the free energy functional
resulting from generalized  Flory equation of state and 
was extended to the case of inhomogeneous solutions of star polymers
\cite{Forsman:04:2}.  However, a very convenient
from numerical point of view approach was developed
by Yu and Wu \cite{Yu}. This approach is based on Rosenfelds'
fundamental measure theory (FMT) \cite{Rosenfeld} and on
Wertheim's first-order thermodynamic perturbation theory (TPT1).
The theory of Yu and Wu
allows for performing quite complex studies because it does not require
single-chain Monte Carlo simulations and yields a fully analytical equation
of state. This approach \cite{Yu} has been successfully applied to investigate
adsorption, surface phase transitions and capillary condensation
in systems involving chain particles
\cite{Bryk,Bryk1,Bryk2}. It was also extended to the case of inhomogeneous
semiflexible and cyclic polyatomic fluids \cite{Yu1}, as well as to
binary hard--rod-polymer mixtures \cite{Bryk03}.

A few years ago Blas and Vega \cite{Blas} proposed an extension of the
associating fluid theory for bulk systems involving branched
chain molecules. 
According to their model, branched
molecules are built of chain segments (arms) of tangentially bonded hard spheres connected via articulation
vertices, each of them formed by $f$ arms. The excess Helmholtz free energy
due to the chain connectivity is separated into two contributions, one
accounting for the formation of the articulation vertex,
 and a second one due
to the formation of the arms. The first term has been described by the
second-order thermodynamic perturbation theory (TPT2), whereas the formation
of chain arms -- via TPT1. 
%
The principal aim of this work is to generalize the bulk theory of
Blas and Vega to the case of inhomogeneous systems. We consider the simplest case
of molecules with one articulation vertex.
The generalization is carried out by utilizing the formalism of Yu and Wu \cite{Yu},
derived for 
chain polymers.

We consider an inhomogeneous fluid composed of star molecules, i.e. each molecule is built of a spherical
``head'' (articulation vertex), and $f$ arms tangentially attached to it.
Each arm is just a chain of $M_f$ tangentially jointed segments. Although the
numbers $M_f$ can be different, in this note we study the case in which
all the arms are of the same length, $M\equiv M_f$, so that
the total number of segments within a molecule is $N=f\cdot M+1$.
All the segments are hard spheres of
diameter $\sigma$. The bonding potential $V_b(\RR)$ is defined so that
$g(\RR)= \exp [-\beta V_b(\RR)]$ is
\begin{equation}\label{eq:1}
g(\RR)=\prod_{i=1}^f\frac{\delta(|\rr_0-\rr^{(i)}_1|-\sigma )}{4\pi
\sigma^2}\nonumber\prod_{j=1}^{M-1}\frac{\delta(|\rr^{(i)}_{j+1}-\rr^{(i)}_j|-\sigma
)}{4\pi \sigma^2},
\end{equation}
where $\mathbf{R}=(\mathbf{r}_0,\{\mathbf{r}^{(i)}_j\})$ with
$i=1,2,\cdots,f$ and $j=1,2,\cdots,M$ denotes the set of segment
positions. The articulation vertex is labelled by the subscript $0$.
All remaining segments are labelled by the superscript 
(specifying arm)
and the 
subscript (specifying the position within a given arm).
The grand potential of the system $\Omega$ is as a functional of the local
density of polymers, $\rho(\RR)$,
\bb\label{eq:2}
\Omega [\rho (\RR)]=F_{id}[\rho(
\RR)]+
F_{ex}[\rho (\RR)]+\int \!\dd%
\RR\rho(\mathbf{R})(V_{ext}(\RR)-\mu),
\ee
where $\mu$ is the con\-fi\-gu\-ra\-tion\-al chemical potential, $V_{ext}$
is the external potential,
$\beta F_{id}[\rho(\RR)]=
\beta \int \!\!\dd\RR\rho (\RR) V_b(\RR)+
\int \!\!\dd\RR\rho (\RR)[\ln (\rho (\RR))-1] $
is the ideal part of the free energy and $F_{ex}$
is the excess free energy. The external potential
is a sum of the potentials acting on each segment, $V_{ext}(\RR)=
v_0(\rr_0)+\sum_{i=1}^f\sum_{j=1}^M v_{j}^{(i)}(\rr_{j}^{(i)})$.
We further assume that the excess free energy
is a functional of the average segment local density defined as
\begin{eqnarray}\label{eq:3}
\rho_s(\rr)&=&
\rho_0(\rr)+
\sum_{i=1}^f\sum_{j=1}^M\rho_j^{(i)}(\rr)=\int\dR\delta(\rr-\rr_0)\rho(\RR)\nonumber\\
&&+\sum_{i=1}^f\sum_{j=1}^M
\int\dR\delta(\rr-\rr_j^{(i)})\rho(\RR),
\end{eqnarray}
where $\rho_j^{(i)}(\rr)$ and $\rho_0(\rr)$ are local densities
of segment ``$j$ within the arm $i$'' and of the articulation segment,
respectively.

Following Yu and Wu \cite{Yu} we decompose the excess free energy 
as
\bb\label{eq:4}
\beta F_{ex}[\rho_s(\rr)]=
\int\dr\left[\Phi_{HS}(\{n_\alpha(\rr)\})+\Phi_{C}(\{n_\alpha(\rr)\})\right]\,,
\ee
where $\Phi_{HS}$ results from the hard-sphere repulsion between segments,
and $\Phi_{C}$ is the contribution due to the connectivity. Each of these
contributions is a function of four scalar and two
vector weighted densities \cite{Yu,Rosenfeld}.
For the hard sphere contribution $\Phi_{HS}$ we use the
White-Bear theory, see Refs.~\cite{ Roth, Yu:02:2}
for the explicit formula.

Wertheim's perturbation theory for a bulk fluid \cite{Wertheim} can be
naturally incorporated into the DFT framework \cite{Chapman}.
The generalization for inhomogeneous star polymer systems is
represented by the expression \cite{Blas,Muller}
\bb\label{eq:5}
\Phi_C(\{n_{\alpha}(\rr)\})=
\Phi^{arm}(\{n_{\alpha}(\rr)\})+\Phi^{art}(\{n_{\alpha}(\rr)\})\;,
\ee
where $\Phi^{arm}$ and $\Phi^{art}$ are the contributions due to the
formation of chains within consecutive arms and due to the formation of
the articulation vertex. The equation for $\Phi^{arm}$ follows
from the theory of Yu and Wu \cite{Yu}
\begin{equation}\label{eq:6}
\Phi^{arm}(\{n_{\alpha}(\rr)\})
=\frac{1+f-N}{N}n_0\zeta\ln [y_{HS}(\sigma,\{n_{\alpha}(\rr)\})],
\end{equation}
where  $\zeta =1-\mathbf{n}_{V2}\cdot \mathbf{n}%
_{V2}/(n_2)^2$, and the contact value of the hard-sphere cavity
function, $y_{HS}(\sigma)$,  results from the Carnahan-Starling
equation of state, cf. Eq.(18) of Ref. \cite{Yu}.
Free energy density  $\Phi^{art}$ is obtained by generalizing
the theory of Blas and Vega \cite{Blas,Muller} 
\bb\label{eq:7}
\Phi^{art}(\{n_{\alpha}(\rr)\})=
\Phi^{art}_{TPT1}(\{n_{\alpha}(\rr)\})+\Phi^{art}_{TPT2}(\{n_{\alpha}(\rr)\})\;,
\ee
where $\Phi^{art}_{TPT1}$ and $\Phi^{art}_{TPT2}$ represent the
first
\bb\label{eq:8}
\Phi^{art}_{TPT1}(\{n_{\alpha}(\rr)\})=-\frac{f}{N}n_0\zeta\ln
[y_{HS}(\sigma,\{n_{\alpha}(\rr)\})]\;,
\ee
and the second-order perturbation terms \cite{Blas,Wertheim}
\begin{eqnarray}\label{eq:9}
\Phi^{art}_{TPT2}&=&\ln\sqrt{1+4\Lambda}\\
&&-\ln\frac{(1+\sqrt{1+4\Lambda})^{f+1}-(1-\sqrt{1+4\Lambda})^{f+1}}{2^{f+1}}\;.
\nonumber
\end{eqnarray}
In the above $\Lambda$ depends on the number of arms and its evaluation
requires the knowledge of the $f$-body correlation function for
$f$ spheres in contact. In the case of $f=3$ the application of the
superposition approximation yields
$\Lambda={(1+a\,n_3+b\,n_3^2)}/{(1-n_3)^3}-1$,
where $a$ and $b$ are constant that depend on the
angles between the arms attached to the articulation vertex
\cite{Blas,Wertheim, Muller}.
In the case of bulk fluids $n_3$ is just the packing fraction. Approximation proposed
for inhomogeneous system relies on substitution of the bulk packing fraction 
by the weighted density $n_3$. Note that this approximation is not unique. 
One can follow the ideas of Yu and Wu\cite{Yu:02:2} and propose an approximation
involving, besides scalar, also vector weighted densities. In this work, however,
we decided to apply as simple expression, as possible. 

\begin{figure}
\includegraphics[clip,width=8cm]{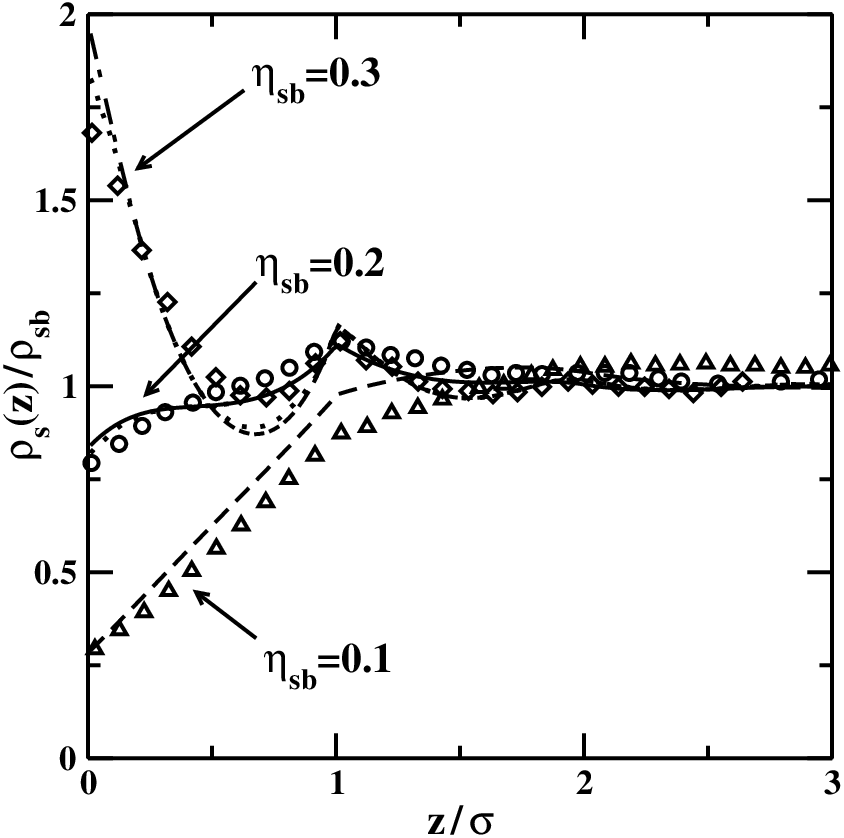}
\caption{\label{fig:1}
The 
average segment density profiles of star polymers ($f=3$, $M=5$)
evaluated for the bulk segment packing fraction
$\eta_{sb}=0.1, 0.2$ and $0.3$. 
Symbols represent computer simulations 
\cite{simul}, dashed lines denote DFT results obtained using TPT1 and
dots -- DFT results evaluated using TPT1 and TPT2 contributions.
}
\end{figure}

Within the TPT1 approach the bulk thermodynamic
properties of the star polymers  are the same as the properties of
chains built of the same number of segments. This is because the first-order
perturbation free energy takes into account only the number of segment
connections and neglects polymer's topology. The latter
is included within the TPT2 approach, cf. Eq.(15) of Ref. \cite{Blas}.
However, the identity of the bulk thermodynamic properties within
the TPT1 theory does not imply that the structure of inhomogeneous
fluids of star polymers and of chains with the same number of segments,
that results from DFT, is identical.

Minimization of (\ref{eq:2}) yields 
\begin{equation}\label{eq:10}
\rho(\RR)=
g(\RR)\exp\left[ \beta\mu -\beta\lambda_0(\rr_0)-
\beta\sum_{i=1}^f\sum_{j=1}^M\lambda_j^{(i)}(\rr_j^{(i)})\right],\,
\end{equation}
where 
$\lambda_j^{(i)}(\mathbf{r}_j^{(i)})={\delta F_{ex}}/{\delta \rho _s(%
\mathbf{r}_j^{(i)})}+v_j^{(i)}(\mathbf{r}_j^{(i)})$ and
$\lambda_0(\mathbf{r}_0)={\delta F_{ex}}/{\delta \rho _s(%
\mathbf{r}_0)}+v_0(\mathbf{r}_0)$.
For systems with the density distribution varying only in the $z$
direction Eqs.~(\ref{eq:3}) and (\ref{eq:10}) lead to the following expressions
for the segment density profiles
\begin{equation}
\rho_0(z_0)=\exp(\beta\mu)\gamma_0(z_0)
\left[G^{M+1}(z_0)\right]^f
\end{equation}
and
\begin{equation}
\rho_j^{(i)}(z_j^{(i)}) = \exp(\beta\mu) 
\gamma_j^{(i)}(z_j^{(i)}) G^{M+1-j}(z_j^{(i)})
\tilde G^{j+1}(z_j^{(i)})\;,
\end{equation}
where $\gamma^{(i)}_j(z)=\exp[-\beta\lambda_j^{(i)}(z)]$;
$\gamma_0(z)\equiv\gamma^{(i)}_0(z)$.
The functions $G^i(z)$
are defined by recurrence relation \cite{Yu}
\begin{equation}
{G}^j(z)=\int\dz'\gamma^{(i)}_j(z')
\frac{\theta(\sigma-|z-z'|)}{2\sigma}{G}^{j-1}(z')\,,
\end{equation}
for $j=2,\cdots,M$ with $G^i(z)\equiv 1$. In the above $\theta$ is the unit-step
function.  The functions $\tilde G^j(z)$, however, are given by
\begin{equation}
\tilde{G}^2(z)=\int\dz'\gamma_0(z') \frac{\theta(\sigma-|z-z'|)}{2\sigma}[G^{M+1}(z')]^{f-1},
\end{equation}
for $j=2$ and
\begin{equation}
\tilde{G}^j(z)=\int\dz'\gamma^{(i)}_j(z')]
\frac{\theta(\sigma-|z-z'|)}{2\sigma}{\tilde G}^{j-1}(z')\,,
\end{equation}
for $j>2$.
The equations given above are valid for the stars with arms
of identical length. In such a case the profiles $\rho_j^{(i)}(z)$ are
independent of the arm index $i$. A generalization of the theory to the case
of stars with arms of different length is straightforward. For example, the
integrand in the last equation would involve a product of the functions
$G^{M_1+1}(z')G^{M_2+1}(z')\dots$, instead of
$[G^{M+1}(z')]^{f-1}$ (here $M_i$'s abbreviate the number of segments within
consecutive arms).
As a simple application of the 
theory we calculate density profiles
of star molecules built of hard-sphere segments near a hard wall, located at
$z=0$. The solutions of the density profile equations were
obtained by using the standard iterational procedure.

\begin{figure}
\includegraphics[clip,width=8cm]{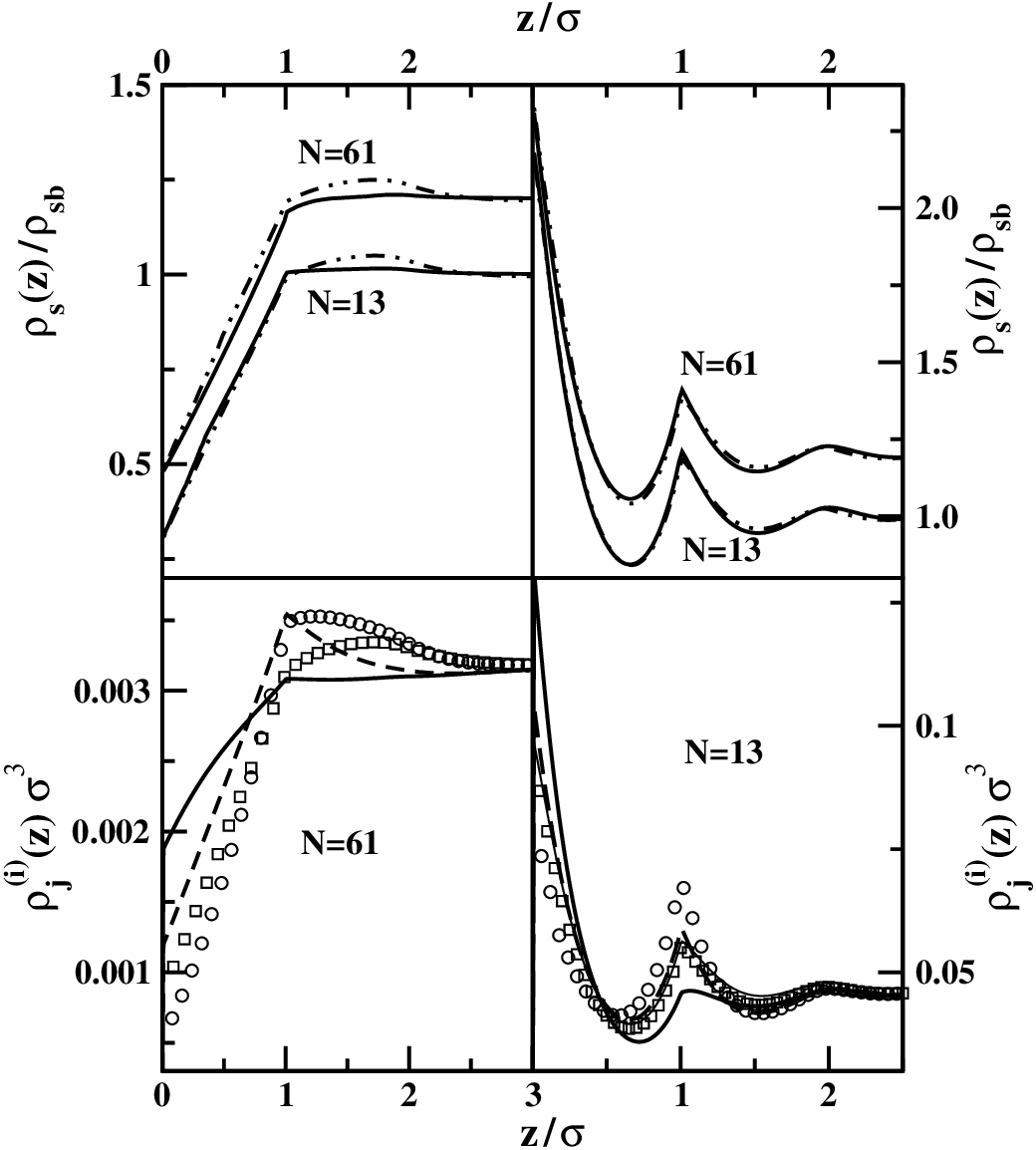}
\caption{\label{fig:2} 
Upper panels:
The 
average segment density profiles of star polymers (dash-dotted lines) and
of chains 
(solid lines) calculated
for the total number of segments $N=13$ and $N=61$. 
The results for $N=61$ are
shifted up by $0.2$. 
Lower panels: 
The 
segment density profiles of the
articulation segment (circles) and of its nearest neighbor (squares) of star
polymers 
and of the first (solid lines)
and the second (dashed lines) segment of 
chains 
In left panels $\rho_{sb}^*=0.2$, whereas $\rho_{sb}^*=0.6$ in right panels.
}
\end{figure}

In Fig.1 we compare the average segment density profiles resulting from 
theory with computer simulations \cite{simul} for star
polymers built of $f=3$ arms, each  composed of $M=5$ segments. The
calculations were carried out for bulk segment packing fractions
$\eta_{sb}=\pi\rho_{sb}\sigma^3/6=0.1, 0.2$ and $0.3$, where $\rho_{sb}$
is the bulk average segment density. The density profiles
in Fig.~1
have been normalized by the bulk density $\rho_{sb}$.
For $\eta_{sb}=0.2$ and 0.3 we show two sets of the DFT results.
The first one has been evaluated employing the TPT1 approach
(i.e. the term given in Eq.~\ref{eq:9} has been neglected), whereas the second
set was obtained using TPT1 and TPT2 contributions
to the free energy. The differences between the local densities
resulting from these two approximations are small and occur only within
the region adjacent to the wall. The TPT2 contribution leads to smaller
contact values of the average segment local density. This effect is quite
obvious, because the TPT2 correction lowers the pressure. The agreement between
theoretical predictions and computer simulations is reasonable.
\begin{figure}
\includegraphics[clip,width=8cm]{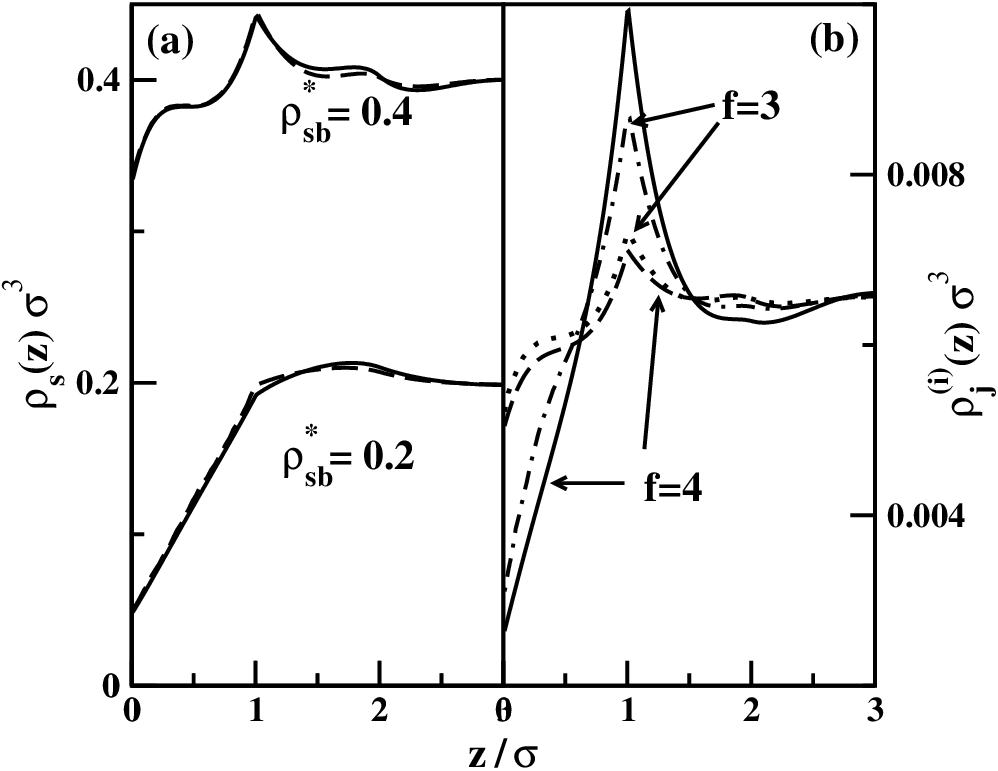}
\caption{\label{fig:3}
(a) The average segment density of star polymers with $f=3$ and 
$M=20$ (dashed line) and with $f=4$ and $M=15$ (solid line) for two bulk
densities, $\rho_{sb}^*=0.2$ and $\rho_{sb}^*=0.4$. (b)
The segment density profiles of ``heads'' (solid and dash-dotted
lines) and it's first neighbor (dashed and dotted lines) for the same
models. The bulk segment density is $\rho_{sb}=0.4$.
}
\end{figure}

Figure 2 compares the density profiles for three-armed star polymers (resulting
from the TPT1 approach) with the profiles of chain polymers built of
the same number of segments obtained from the approach of Yu and Wu \cite{Yu}.
The results are for bulk average segment densities
$\rho_{sb}^*=\rho_{sb}\sigma^3=0.2$ and $0.6$ and for two model systems
with different total number of segments $N=13$ and $N=61$.
For $N=13$ each star polymer arm is composed of $M=4$ segments, whereas
for $N=61$ each arm consists of $M=20$ segments. In the upper panels
we compare the average segment densities normalized to unity. We find that for
higher bulk density $\rho_{sb}^*=0.6$ (the upper right-hand side panel)
the local densities $\rho_s(z)$ of chains and stars are quite similar.
The profiles are dominated by packing effects. Larger differences between
the star and chain polymer profiles are visible at lower density,
$\rho_{sb}^*=0.2$, cf. the upper left-hand side panel.
Note that the contact values of $\rho_s(z)$ for star and chain polymers are
identical in our TPT1 approach.
Lower panels of Fig. 2 show the density profiles of selected segments for
the same systems. We plot here the profiles of ``heads'' (in the case of
chains the profiles of the first segment) and the profiles of the segments
attached directly to the ``heads''. The differences between the profiles for
the chain and star polymers are now more pronounced, especially for the
profiles of the ``heads''. We have also inspected the profiles for the
segments that are topologically more distanced from the head and have found
that the difference between them becomes gradually smaller.

Finally Fig. 3 presents the profiles of the stars built of the same number of
segments, but having different number of arms.  We have considered the
models with $f=3$, $M=20$ and with $f=4$, $M=15$. It is not surprising
that the difference between the average segment density profiles (cf. Fig.~3a)
is now less pronounced than in the case of the profiles
displayed in Fig. 2, because the differences in the topology of the particles
are now smaller. However, the differences between the individual
segment density profiles still persist, cf. Fig.~3b, where we show
the profiles of ``heads'' and the segments directly attached
to heads.

In conclusion, in this work we propose density functional theory
for inhomogeneous star polymers. Although the theory is written down for
the case of arms composed of identical
numbers of segments, its generalization for stars with arms of different length
is straightforward. 
Several further extensions
are also possible. In particular it would be of interest to consider cases of
physically different ``head'' and ``arm'' segments in order to mimic the
systems involving surfactants. Some of these topics are already under study in
our laboratory.

This work was supported by EU under  a TOK contract No. 509249.
We are grateful to prof. O. Pizio for early discussions and bringing
Ref.~\cite{Blas} to our attention.


\begin{thebibliography}{99}

\bibitem{Likos} C. N. Likos, Phys. Rep. {\bf 348}, 247 (2001).

\bibitem{Mayer} C. Mayer, C. N. Likos, and H. L\"owen, Phys. Rev. E
\textbf{70}, 041402 (2004).

\bibitem{Woodward91} C.E. Woodward, J. Chem. Phys. {\bf 94}, 3138 (1991).

\bibitem{Yeti95} A. Yethiraj and C.E. Woodward, J. Chem. Phys. {\bf 102},
5499 (1995).

\bibitem{Yeti} A. Yethiraj, J. Chem. Phys. \textbf{109}, 3269 (1998).

\bibitem{Yu}  Y.-X. Yu and J. Wu, J. Chem. Phys. \textbf{117}, 2368
(2002).

\bibitem{Louis00} A.A. Louis, P.G. Bolhuis, J.P. Hansen, and E.J. Meijer,
Phys. Rev. Lett.  {\bf 85}, 2522 (2000).


\bibitem{Archer02} A.J. Archer, C.N. Likos, and R. Evans,
J. Phys.: Condens. Matter {\bf 14} 12031 (2002).

\bibitem{Yu1} D. Cao and J. Wu, J. Chem. Phys. \textbf{121}, 4210 (2004).

\bibitem{Bryk04} P. Bryk and S. Soko\l owski, J. Chem. Phys.
\textbf{120}, 8299 (2004).

\bibitem{Forsman:02:1} J. Forsman, C.E. Woodward and B. C. Freasier, J. Chem. Phys,
{\bf 117}, 1915 (2002).

\bibitem{Forsman:04:2} C. E. Woodward and J. Forsman, Macromolecules, {\bf 37}, 7034 (2004).

\bibitem{Rosenfeld}  Y. Rosenfeld, Phys. Rev. Lett. \textbf{63}, 980
(1989).

\bibitem{Bryk} P. Bryk and S. Soko\l owski, J. Chem. Phys. \textbf{121},
11314 (2004).

\bibitem{Bryk1} P. Bryk, K. Bucior, S. Soko\l owski, and G. \.Zukoci\'nski,
J. Phys. Chem. B \textbf{109}, 2977 (2005).

\bibitem{Bryk2} P. Bryk, O. Pizio, and S. Soko\l owski, J. Chem. Phys. 
\textbf{122}, in press (2005).

\bibitem{Bryk03} P. Bryk, Phys. Rev. E
\textbf{68}, 062501 (2003).

\bibitem{Blas} F.J. Blas and L.F. Vega, J. Chem. Phys. \textbf{115}, 3906
(2001).

\bibitem{Roth}  R. Roth, R. Evans, A. Lang, and G. Kahl, J. Phys.:
Condens. Matter \textbf{14}, 12063 (2002).
\bibitem{Yu:02:2}  Y.-X. Yu and J. Wu, J. Chem. Phys. \textbf{117}, 10156
(2002).

\bibitem{Chapman} W.G. Chapman, Ph.D. Thesis Cornell University, Ithaca
(1988).
\bibitem{Wertheim}  M.S. Wertheim, J. Chem. Phys. \textbf{87}, 7323
(1987).
\bibitem{Muller} E.A. M\"uller and K.E. Gubbins, Mol. Phys. \textbf{80}, 957
(1993).
\bibitem{simul} A. Yethiraj and C.K. Hall, J. Chem. Phys. \textbf{94}, 3943
(1991).

\end{thebibliography}
\end{document}